# Comment on
# Quantum Interference between Light Sources Separated by 150 Million Kilometers
# Deng et al. PHYSICAL REVIEW LETTERS 123, 080401 (2019)

*Shaul Mukamel*
*Department of Chemistry, and Physics and Astronomy University of California, Irvine, Irvine, CA 92697*

In the paper of Deng et al, sunlight is attenuated down to the single photon level and the photon is then manipulated to mimic another photon generated by a quantum dot (QD) source, in all degrees of freedom: polarization, spatial modes, spectral and temporal properties, and also photon statistics.

The resulting manipulated-sun-photon (MSP) and the QD photon are then shown to undergo a Hong-Ou-Mandel (HOM) two-photon interference, which reveals that the two photons are indistinguishable quantum particles. This inspiring experiment provides a clear illustration for the quantum nature of single photon states, which is well established in quantum optics, and involves a remarkable application of state-of-the-art quantum optics, detection technologies, and high-performance single-photon sources

On a closer look, the analysis presented in that paper is misleading. The MSP (and hence the HOM signal) contains no information about the sunlight and its statistics. The photon manipulation process has carefully erased all such information. The experiment is thus an impressive demonstration of photon fabrication and manipulation ability, but has no relevance to the state of sunlight. Any claim that this study shows that sunlight is somehow quantum ,such as the statement. "Providing unambiguous evidence of the quantum nature of thermal light."  is false. The   following claim is also baseless since the experiment carries no information about sunlight.

 "We expect their coherence properties may reveal dynamic sun activities, for example, dramatic changes in magnetic field, reemission of photons at dark lines, dispersion (chirping) mechanism, etc."

The fact that a sunlight photon happened to be used to create the MSP is insignificant. In the end, what is detected is the MSP, not the original sunlight photon. While the use of quantum interferometric techniques in astronomy, as suggested by the authors, is an intriguing, worth exploring possibility, the present work does not provide any basis for such claim.